\documentclass[11pt,twoside]{article}  
\usepackage{asp2006}
\usepackage{adassconf}

\def\swift{\emph{Swift}}
\def\etal{et al.\ }
\def\til{\ensuremath{\sim}}

\begin{document}   

%

\paperID{P5.3}

%

\title{Automatic analysis of Swift-XRT data}
       
%
%
%
%
%

\markboth{Evans et al.}{Automatic data analysis}

%

\author{P.\ A.\ Evans, L.\ G.\ Tyler, A.\ P.\ Beardmore, J.\ P.\ Osborne}
\affil{Department of Physics and Astronomy, University of Leicester,
Leicester, LE1 7RH, UK}


\contact{Phil Evans}
\email{pae9@star.le.ac.uk}

%
%

\paindex{Evans, P.~A.}
\aindex{Tyler, L.~G.}
\aindex{Beardmore, A.~P.}
\aindex{Osborne, J.~P.}


\keywords{astronomy!astrometry,astronomy!X-ray,data processing!real-time}



\begin{abstract}          
The \swift\ spacecraft detects and autonomously observes \til100 Gamma Ray
Bursts (GRBs) per year, \til96\%\ of which are detected by the X-ray telescope
(XRT). GRBs are accompanied by optical transients and the field of ground-based
follow-up of GRBs has expanded significantly over the last few years, with rapid
response instruments capable of responding to \swift\ triggers on timescales of
minutes. To make the most efficient use of limited telescope time, follow-up
astronomers need accurate positions of GRBs as soon as possible after the
trigger. Additionally, information such as the X-ray light curve, is of interest
when considering observing strategy. The \swift\ team at Leicester University
have developed techniques to improve the accuracy of the GRB positions available
from the XRT, and to produce science-grade X-ray light curves of GRBs. These
techniques are fully automated, and are executed as soon as data are available.
\end{abstract}


\section{Introduction}
The \swift\ satellite triggers on \til100 Gamma Ray
Bursts (GRBs) per year, and the X-ray telescope (XRT, Burrows \etal2005)
provides localisations accurate to \til5 arcsec for $>90\%$ of these. However,
GRBs, and their optical counterparts, fade rapidly and ground-based observers
have limited telescope time available to use for observations. It is thus
desirable to produce precise, accurate positions rapidly. Furthermore, a key
indication of the scientific interest of a GRB comes from the \swift-XRT light
curve. These are non-trivial to produce correctly, but it is desirable to
generate them accurately and rapidly.

We describe how these two challenges are being met by the \swift\ team at
Leicester, providing automatic data analysis which gives results of
publication-grade quality. While our work is specific to \swift, rapid response
time-domain astrophysics is a fast-growing field, and this type of software will
be increasingly useful to astronomers.

\section{Positions}
To locate sources on the XRT detector we first perform a cell-detect search. We
then centroid accurately on these sources using the point spread function
(PSF)-fitting technique described by Cash (1978). Since the XRT's PSF is known,
our fit has just two interesting free parameters, the $x$ and $y$ position of
the object. Further, the presence of hot pixels has a very minor effect on the
fit and will be correctly accounted for in the uncertainty. Following a
micrometroid impact in mid-2005, there area several bad columns on the XRT
detector, which are permanently masked out. The location of these is known, and
the fitted PSF is adjusted to account for this, giving accurate positions even
when the object lies right over the bad columns (Fig.~1).

\subsection{Rapidly available positions}
When \swift\ detects a GRB, the XRT takes up to three short ($<3$ s) images and
attempts to find a source; if successful it performs a barycentric centroid and
dispatches a GCN Notice announcing the position to the community. For the first
spacecraft orbit after a GRB is detected, limited data products are immediately
telemetered to the ground, some of which (`SPER' -- Single Pixel Event Report --
data) can be used to improve this position, or determine a position if none was
found in the short images. As soon as SPER data are received, usually within
10--20 minutes of a GRB detection, an automatic
process triggers the search and centroid routine described above. Within seconds
the source position is found and relayed to the international community via the
GCN Notices system. Results are also published online at
http://www.swift.ac.uk/spertable.php.

\begin{figure}[t]
\epsscale{0.32}
\plottwo{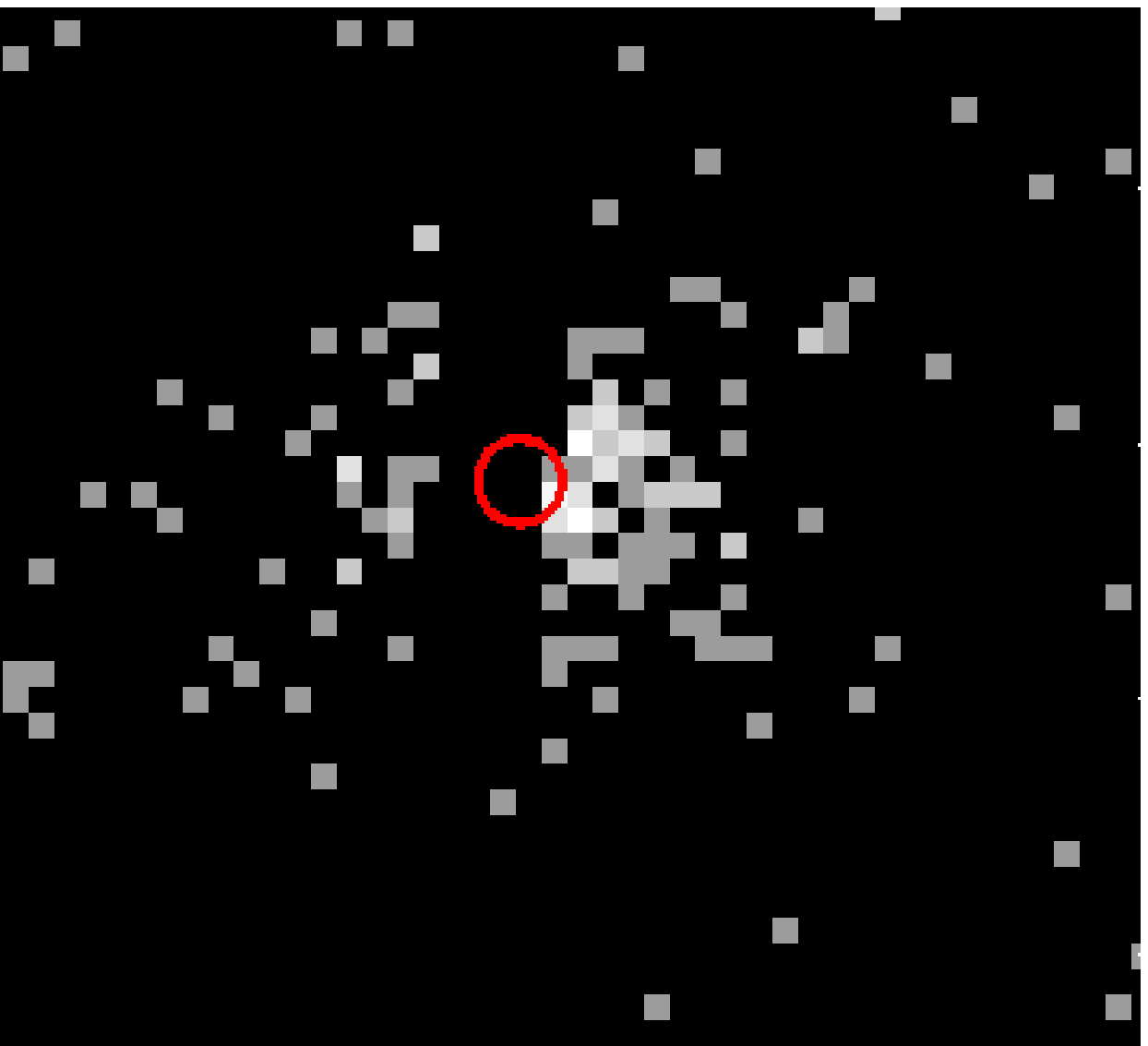}{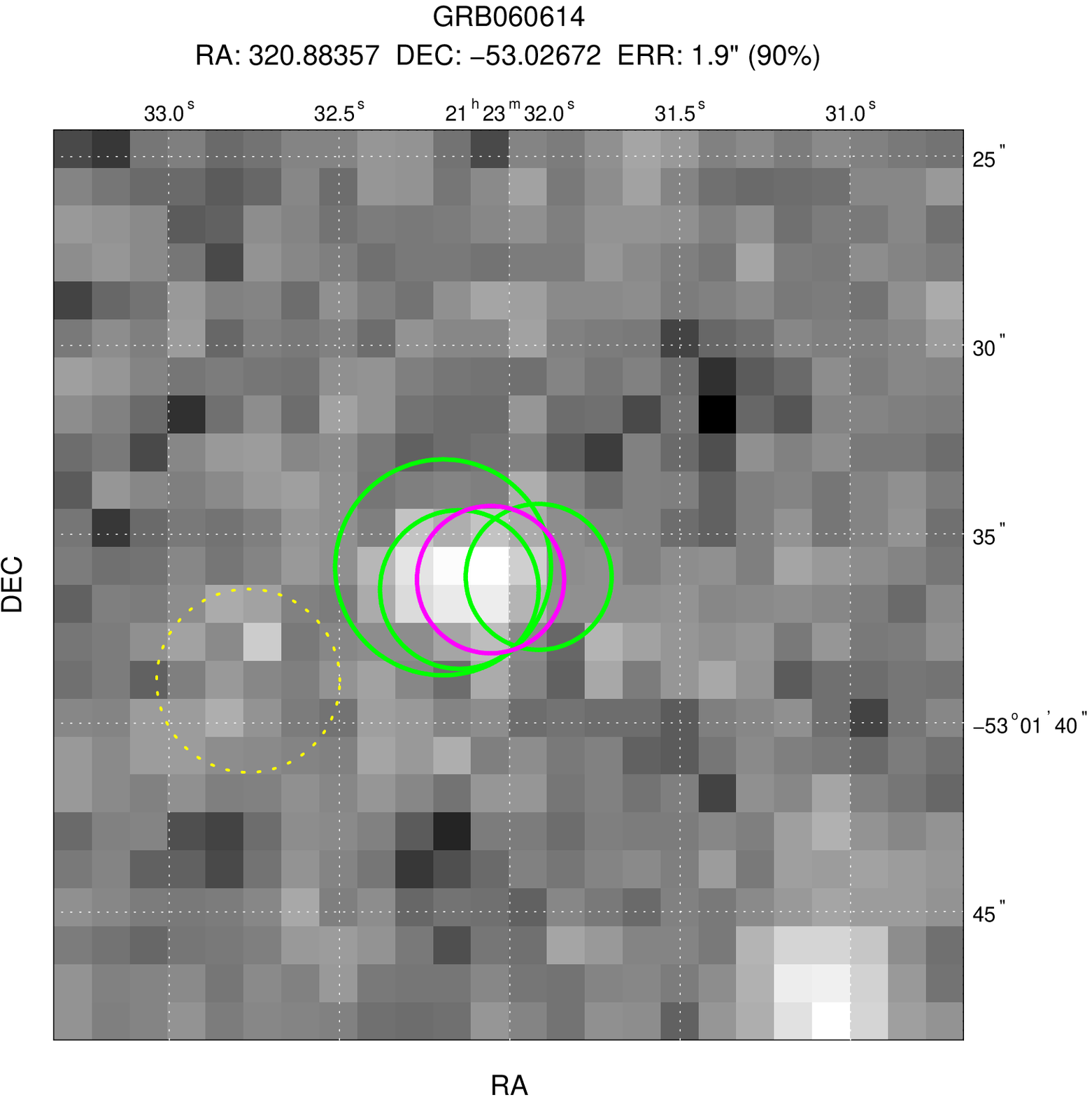}
\caption{\emph{Left}: An image of GRB 070429A, generated from SPER data. The red circle shows the
automatically determined GRB position. Despite the bad columns, the
centroid is accurate.
\newline
\emph{Right}: The UVOT-enhanced XRT position of GRB 060614. The image is the UVOT $V$-band
image. The magenta circle shows the final position. The green circles are the
individual positions which contributed to this, and the yellow circle was
rejected as an outlier. }
\label{fig:sper}
\end{figure}

\subsection{Enhanced positions} The above positions are subject to the
uncertainty in the spacecraft boresight, derived from the on-board star tracker,
which is approximately 3.5". We have developed a technique to remove much of
this uncertainty.

In addition to the XRT, \swift\ contains an ultra-violet and optical telescope
(UVOT, Roming \etal2005) which takes data simultaneously with the XRT. We have deduced the
transformation from a position on the XRT detector to its equivalent on the UVOT
(when the $V$ filter is in use), thus for any source detected in XRT we can
determine where it would appear on the UVOT. We then use the standard \swift\ 
software to translate this into an initial sky position, match serendipitous sources in
the UVOT field of view with the USNO-B1 catalogue to determine an aspect
correction, and apply this correction to find the true sky position. This removes the
spacecraft boresight from the loop entirely, the position accuracy being limited
by the accuracy of the XRT to UVOT transformation and the accuracy of the aspect
solution. \swift\ usually takes multiple observations of GRB, so we have have
multiple datasets on which the above technique can be applied. We can then take
the weighted mean of the positions thus produced which reduces the uncertainty
arising from the aspect solution. Any outliers are detected and removed, and the
weighted mean is recalculated. Fig.~\ref{fig:sper} shows an example GRB with the
weighted mean position, the individual positions which contributed to this, and
an outlier. The 90\% error radii of these final `UVOT-enhanced XRT positions' are
$<1.9$" 50\% the time, a factor of 2 reduction compared to the normal,
unenhanced positions.

This process is fully automated, and runs as soon as \swift\ data of a GRB are
received by the UK Swift Science Data Centre (UKSSDC). Once a position is
determined a GCN Circular is automatically dispatched, advising the community of
the position, which is also posted online at
\mbox{http://www.swift.ac.uk/xrt\_positions}. As more data are received, this
position is continually revised. Further GCN circulars are not produced, however
the website is updated with each run of the software. Full details of this
procedure are described in Goad \etal(2007).

\section{Light curves}
The standard approaches to light curve creation assume essentially uniform event
data and produce uniform bin sizes, however for GRBs, whose essential behaviour
is to fade, this is not appropriate. Further, \swift-XRT data are complicated 
by the dead columns on the CCD, the XRT's innovative mode-switching technology
which allows the XRT to determine its operating mode based on source brightness
(which changes), and pile-up -- where multiple photons arrive on the same XRT
pixel during one readout cycle, and are thus recorded as only a single photon.
It is vital that these effects are properly accounted for: optical astronomers,
who must decide whether to invest their limited telescope time on a given GRB,
must have confidence in light curve features such as gradient changes (`breaks')
or flares, which may guide their decision on whether to observe a given burst.

\begin{figure}[t]
\epsscale{0.40}
\plottwo{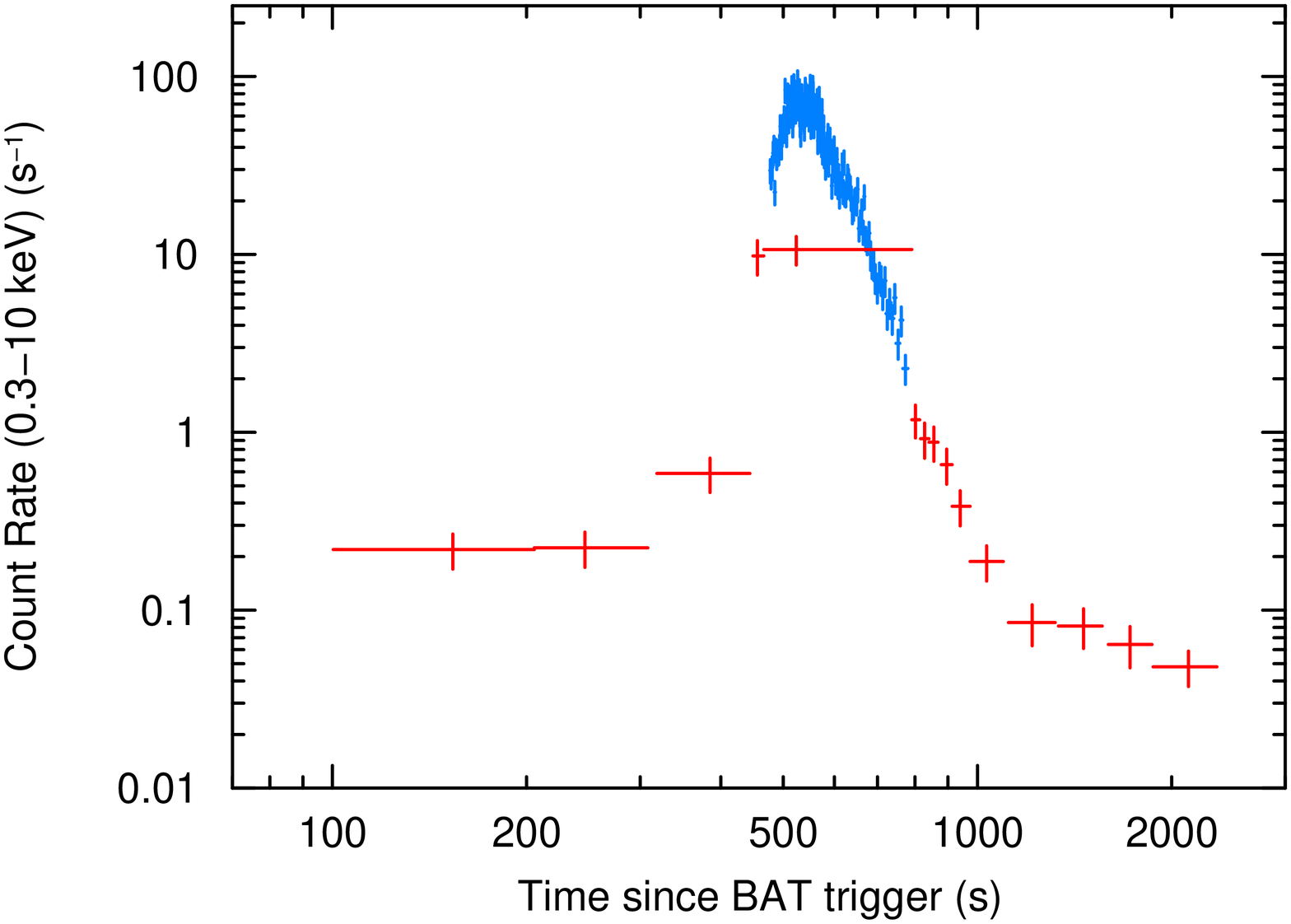}{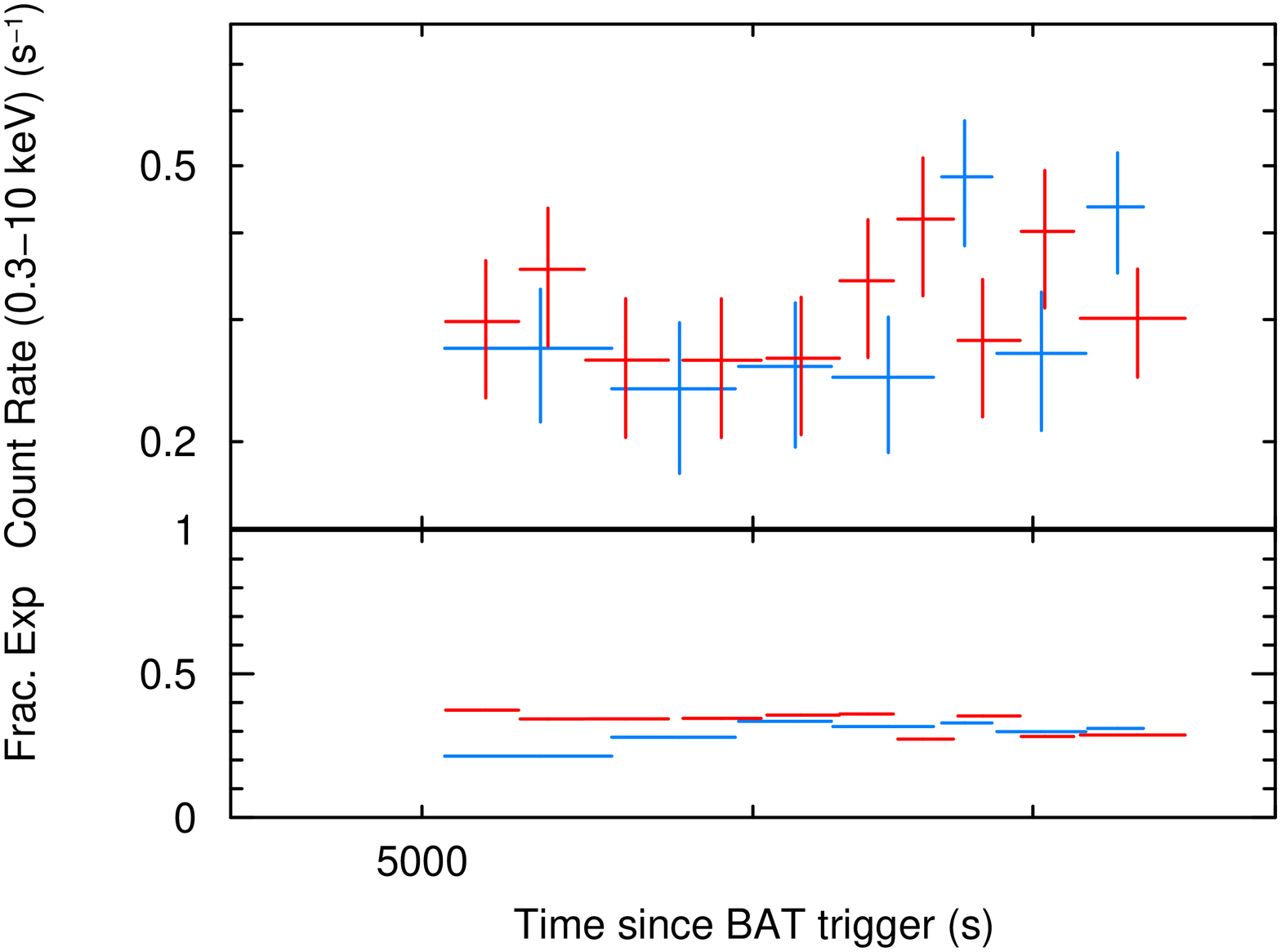}
\caption{Examples of readout-mode switching/ \emph{Left}: GRB 060929, showing a 
giant flare. Where data from the two modes overlap the red data-point aroiund
500 s is spurious, and will be rejected by our code (it is left in here for
illustrative purposes). \emph{Right}: GRB 050315. Here, the
readout mode toggled rapidly, and our software correctly decided that both
datasets are both valid, will keep them.}
\label{fig:ms}
\end{figure}

To this end, we have developed a fully automated script which runs every time
new \swift\ data of GRBs arrive at the UKSSDC, approximately every ninety
minutes. This software dynamically varies the source extraction region and the
data-bin size based on the source count-rate (corrected for losses due to
pile-up) It builds separate light curves for the different operating modes, and
then combines them. Where data from multiple modes overlap, it decides whether
both data points are valid, or one is spurious (Fig.~\ref{fig:ms}). If the
instrument is toggling rapidly between modes, the data from both modes should be
correct. However, if the instrument spends a considerable amount of time in one
mode, but a datapoint from another mode spans this interval, the latter point
will be an average of the count-rate before and after the mode-switch, rather
than during it, so should be rejected.  Our code also models the PSF of the data
to determine whether pile-up was a factor, and if so excludes the bright core of
the source from the data extraction, and then corrects the count-rates for data
thus lost. For any times where source counts were lost due to the bad columns,
the number of lost counts is determined, and the count-rate amended accordingly.

The veracity of these light curves have been extensively tested against those
manually built by experienced members of the XRT team, and they are confirmed to
be reliable and accurate, suitable both for guiding observing strategy and for
use in refereed publications. Once created, light curves are published online at
http://www.swift.ac.uk/xrt\_curves in both graphical (ps, gif) form, and as
ASCII data files. These are updated whenever new data are received.

Full details of this procedure are described in Evans \etal(2007).

\section{Closing Remarks}
The automated science-grade analysis of data is both achievable, and essential
in some fields. With the increasing popularity of time-domain astrophysics, and
the advent of rapid response hardware, the demand for such software will only
increase.


\end{document}